\documentclass[aps,showpacs,tightenlines,twocolumn,nofootinbib,nobibnotes,superscriptaddress]{revtex4-1}
\usepackage{amsmath,amssymb,amsfonts,bm}
\usepackage{graphicx}
\usepackage{epstopdf}
\usepackage{dcolumn}
\usepackage{mathrsfs}
\usepackage{tgtermes}
\usepackage[colorlinks=true,linkcolor=red,citecolor=blue, urlcolor=blue,bookmarks=false]{hyperref}
\bibpunct{[}{]}{,}{n}{}{}
\usepackage{verbatim}

\begin{document}
\title{Chern insulator in a hyperbolic lattice}
\date{\today }
\author{Zheng-Rong Liu}
\author{Chun-Bo Hua}
\author{Tan Peng}
\author{Bin Zhou}\email{binzhou@hubu.edu.cn}
\affiliation{Department of Physics, Hubei University, Wuhan 430062, China}

\begin{abstract}
Motivated by the recent experimental realizations of hyperbolic lattices in circuit quantum electrodynamics and the research interest in the non-Euclidean generalization of topological phenomena, we investigate the Chern insulator phases in a hyperbolic $\{8,3\}$ lattice, which is made from regular octagons ($8$-gons) such that the coordination number of each lattice site is $3$. Based on the conformal projection of the hyperbolic lattice into the Euclidean plane, i.e., the Poincar\'{e} disk model, by calculating the Bott index ($B$) and the two-terminal conductance, we reveal two Chern insulator phases (with $B=1$ and $B=-1$, respectively) accompanied with quantized conductance plateaus in the hyperbolic $\{8,3\}$ lattice. The numerical calculation results of the nonequilibrium local current distribution further confirm that the quantized conductance plateau originates from the chiral edge states and the two Chern insulator phases exhibit opposite chirality. Moreover, we explore the effect of disorder on topological phases in the hyperbolic lattice. It is demonstrated that the chiral edge states of Chern insulators are robust against weak disorder in the hyperbolic lattice. More fascinating is the discovery of disorder-induced topologically nontrivial phases exhibiting chiral edge states in the hyperbolic lattice, realizing a non-Euclidean analog of topological Anderson insulator. Our work provides a route for the exploration of topologically nontrivial states in hyperbolic geometric systems.

\end{abstract}

\maketitle

\section{Introduction}

Recently, a great deal of research effort in condensed matter physics has been made to find novel topological states. A typical example is Chern insulator (CI)~\cite{RevModPhys.82.3045, RevModPhys.83.1057, RevModPhys.88.021004}, which possesses fully gapped bulk states and gapless boundary states and is characterized by a non-zero Chern number. The research branches of CIs include quantum Hall effect (QHE)~\cite{PhysRevLett.45.494, PhysRevLett.48.1559, PhysRevLett.49.405, zhang2005experimental, PhysRevB.62.R6065, PhysRevB.104.115411, PhysRevLett.127.066801, PhysRevResearch.3.033227} and quantum anomalous Hall effect (QAHE)~\cite{PhysRevLett.61.2015, PhysRevB.29.6012, PhysRevB.30.1069, PhysRevLett.50.1395, PhysRevLett.62.1181, PhysRevLett.109.055502, PhysRevB.87.155415, doi:10.1080/00018732.2015.1068524, doi:10.1126/science.1187485, doi:10.1126/science.1234414, doi:10.1146/annurev-conmatphys-031115-011417, PhysRevLett.101.146802, 10.1093/nsr/nwt029, Chang_2016, PhysRevLett.113.147201, PhysRevB.82.115124, PhysRevB.84.195444, PhysRevLett.107.256801, PhysRevB.83.155447, PhysRevB.82.161414, PhysRevLett.106.156801, PhysRevLett.110.026603, PhysRevB.86.035104, PhysRevB.87.205132, PhysRevLett.110.196801, PhysRevLett.110.106804, PhysRevB.95.045113, PhysRevB.97.094408, PhysRevApplied.12.024063, C6NR08522A, D0MH00396D, doi:10.1126/sciadv.aaw5685, PhysRevLett.113.137201, PhysRevLett.114.187201, checkelsky2014trajectory, chang2015high, doi:10.1126/science.aax8156}, which have been extensively investigated in both theory and experiment. Unlike the QHE, the QAHE does not require a strong external magnetic field, which means that it has a unique application potential in electronic devices with low-power consumption~\cite{10.1093/nsr/nwt029, Chang_2016}. The first prediction for the realization of the QAHE was proposed by Haldane in 1988~\cite{PhysRevLett.61.2015}. Although this proposal has been implemented in the ultracold atomic system~\cite{Jotzu_2014}, it is still difficult to achieve in condensed matter systems. The discovery of topological insulators (TIs) provides a good platform for searching the QAHE~\cite{doi:10.1126/science.1133734, doi:10.1126/science.1148047}. The QAHE is predicted to be produced by doping magnetic impurities in HgTe/CdTe quantum wells or InAs/GaSb quantum wells~\cite{PhysRevLett.101.146802, PhysRevLett.113.147201}. Moreover, there are a large number of proposals for realizing the QAHE in various systems, such as single-layer and bilayer graphene~\cite{PhysRevB.82.115124, PhysRevB.84.195444, PhysRevLett.107.256801, PhysRevB.83.155447, PhysRevB.82.161414, PhysRevLett.106.156801}, silicene~\cite{PhysRevLett.109.055502, PhysRevLett.110.026603}, Bi (111) bilayers~\cite{PhysRevB.86.035104, PhysRevB.87.205132}, two-dimensional (2D) organic materials~\cite{PhysRevLett.110.196801, PhysRevLett.110.106804}, transition metal (TM) halides~\cite{PhysRevB.95.045113, PhysRevB.97.094408, PhysRevApplied.12.024063, C6NR08522A, D0MH00396D}, TM atom-doped tetradymite semiconductors~\cite{doi:10.1126/science.1187485}, van der Waals layered MnBi$_{2}$Te$_{4}$-family materials~\cite{doi:10.1126/sciadv.aaw5685}. The QAHE was first observed experimentally in the Cr-doped (Bi, Sb)$_{2}$Te$_{3}$ thin film~\cite{doi:10.1126/science.1234414}. Subsequently, there have been many experiments to study the QAHE in Cr-doped (Bi, Sb)$_{2}$Te$_{3}$~\cite{PhysRevLett.113.137201, PhysRevLett.114.187201, checkelsky2014trajectory} and V-doped (Bi, Sb)$_{2}$Te$_{3}$~\cite{chang2015high}. Recently, the QAHE has been observed in the layered magnetic TI  MnBi$_{2}$Te$_{4}$~\cite{doi:10.1126/science.aax8156} and a MnBi$_{2}$Te$_{4}$/Bi$_{2}$Te$_{3}$ superlattice~\cite{Deng_2020}.

Up to now, most of the research in CIs and other topological states are associated with Euclidean geometry. Due to the intriguing properties of non-Euclidean geometry, the non-Euclidean generalization of topological phenomena is a fascinating issue. Recently, Yu $et~al$. theoretically explored the topological quantum phenomenon in 2D hyperbolic lattices and demonstrated a non-Euclidean analog of the quantum spin Hall effect~\cite{PhysRevLett.125.053901}. Hyperbolic lattices are an important platform for quantum simulation in a curved space with the constant negative curvature, and also have exciting applications in the field of fault-tolerant quantum computing~\cite{PhysRevA.102.032208, 7456305, Breuckmann_2017, Lavasani2019universallogical, Jahn_2021, pastawski2015holographic}. In contrast to Euclidean lattices, hyperbolic lattices are embedded in the hyperboloid and cannot be realized in Euclidean space without distortion. Notably, recent experimental realization of hyperbolic lattices in circuit quantum electrodynamics has been reported \cite{Koll_r_2019}. This experimental breakthrough inspired several subsequent theoretical studies in hyperbolic lattices \cite{PhysRevLett.125.053901, bienias2021circuit, PhysRevA.102.032208,doi:10.1126/sciadv.abe9170,Ikeda_2021,boettcher2021crystallography, PhysRevLett.128.166402}. In our work, motivated by the recent experimental realization of hyperbolic lattices and theoretical progress in topological hyperbolic lattices, we aim to extend CI phases and consequent topological phenomena to non-Euclidean hyperbolic geometry. Due to the absence of commutative translation symmetries and Bravais vectors in hyperbolic lattices, we will apply exact numerical diagonalization of the Hamiltonian to obtain the energy spectrum and adopt the real-space Bott index to characterize the topological phase in a hyperbolic lattice. Further, we numerically calculate the two-terminal conductance and the nonequilibrium local current distribution to analyze the topological phenomena of CI in the hyperbolic lattice.

On the other hand, the interplay between topology and disorder plays an important role in the recent research of topological states with Euclidean geometry. Li $et~al$. found that strong disorder can destroy the topological insulate state, and a certain strength of disorder can induce the system to transform from a normal insulator to topologically nontrivial insulator called topological Anderson insulator (TAI)~\cite{PhysRevLett.102.136806}. After Li $et~al$. first proposed this concept~\cite{PhysRevLett.102.136806}, TAI and disorder effects have been investigated in various systems~\cite{PhysRevB.80.165316, PhysRevLett.103.196805, PhysRevLett.105.216601, PhysRevB.84.035110, PhysRevB.85.035107, PhysRevB.85.195125, PhysRevB.85.195140, doi:10.1063/1.4829683, PhysRevLett.112.206602, PhysRevB.91.214204, PhysRevB.93.214206, Orth_2016, PhysRevB.100.184202, PhysRevLett.125.133603, PhysRevB.95.245305, PhysRevB.100.115311, PhysRevB.103.085408, PhysRevB.103.085307, PhysRevB.104.245302, PhysRevB.103.224203}, including HgTe/CdTe quantum wells~\cite{PhysRevB.80.165316}, Kane-Mele models~\cite{Orth_2016}, electronic circuits~\cite{PhysRevB.100.184202}, photonic crystals~\cite{PhysRevLett.125.133603}, quasicrystals~\cite{PhysRevB.100.115311, PhysRevB.103.085307, PhysRevB.104.245302}. In experiments, TAIs have been successively realized in a photonic platform~\cite{stutzer2018photonic}, an atomic wire~\cite{doi:10.1126/science.aat3406}, and electronic circuits~\cite{PhysRevLett.126.146802}. In the recent research of topological phases of matter in hyperbolic geometry, the disorder-immune behaviors of the topologically protected edge states in topological hyperbolic lattices have been demonstrated~\cite{PhysRevLett.125.053901}. However, the investigation of the disorder-induced topological phases in non-Euclidean hyperbolic lattices has yet to be reported. An interesting question is whether the TAI can be realized in hyperbolic lattices. Thus, in our work, we also explore the effect of disorder on topological phases in the hyperbolic lattice. It is demonstrated that the chiral edge states of CIs are robust against weak disorder in the hyperbolic lattice. More striking, the disorder-induced topologically nontrivial phases exhibiting chiral edge states are found at a certain region of disorder strength in the hyperbolic lattice, which is a non-Euclidean extension of TAI.

The rest of the paper is organized as follows. In Sec.~\ref{Model}, we introduce a CI model in the hyperbolic lattice. Then, in the clean limit, we investigate the topological phase transition and transport properties of CI in the hyperbolic lattice through numerical calculations in Sec.~\ref{Numerical Results} (A). Next, we explore the effect of disorder on CI in the hyperbolic lattice in Sec.~\ref{Numerical Results} (B). Finally, we summarize our conclusions in Sec.~\ref{Conclusion}.

\section{Models}
\label{Model}

The Schl\"{a}fli symbol $\{p, q\}$ is used to describe the essential characteristics of regular polygons, where $p$ and $q$ represent the number of vertices of the regular polygon and the number of polygons adjacent to each vertex, respectively~\cite{coxeter1973regular}. In Euclidean geometry, the Gaussian curvature of the surface is equal to $0$, and $\{p, q\}$ satisfies the relation $(p-2)\cdot(q-2)=4$. In the Euclidean plane, there are only three pairs $\{p=3, q=6\}$, $\{p=4, q=4\}$ and $\{p=6, q=3\}$ meeting the conditions, which means that only lattices with 3-fold, 4-fold, and 6-fold rotational symmetry can be periodically tessellated in 2D Euclidean space~\cite{10.2307/3647934}. However, in hyperbolic geometry, the Gaussian curvature of the surface is less than $0$, and $\{p, q\}$ satisfies the relation $(p-2)\cdot(q-2)>4$. It means that the hyperbolic lattice holds infinite possibilities of regular tessellation in the hyperbolic space. The Poincar\'{e} disk model, which projects the entire hyperbolic plane into a unit circle in the Euclidean plane, provides a visual approach to understand hyperbolic geometry~\cite{weeks2001shape, greenberg1993euclidean}.

\begin{figure}[t]
	\includegraphics[width=7cm]{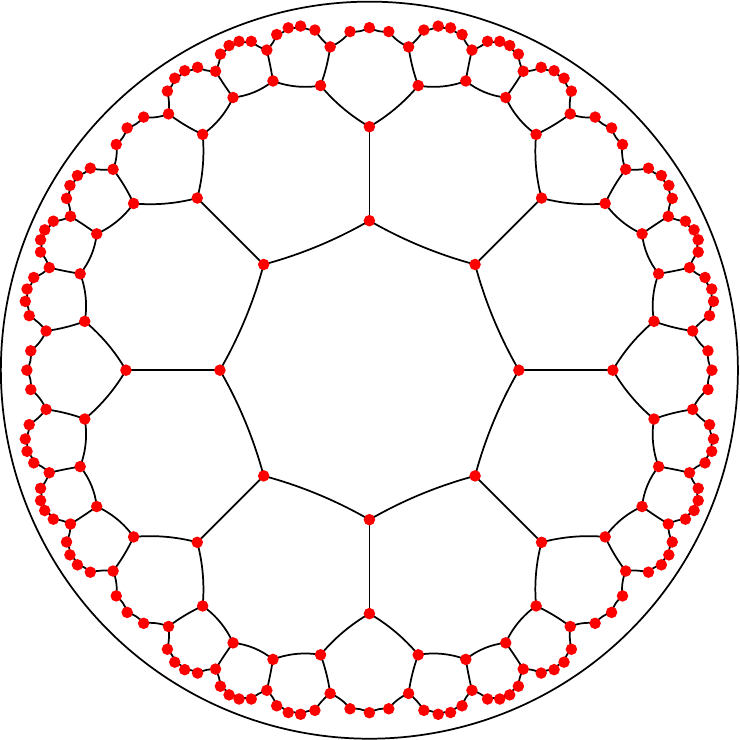} \caption{Schematic illustration of the hyperbolic $\{8,3\}$ lattice at the epoch $3$. In the Poincar\'{e} disk, the vertices of these polygons are the hyperbolic lattice sites, and the non-Euclidean metric is such that all neighboring lattice sites have equal hyperbolic distance.}%
	\label{fig1}
\end{figure}

In this work, in order to investigate the topological properties of hyperbolic geometric systems, we focus on the hyperbolic $\{8,3\}$ lattice for concreteness, which corresponds to a tessellation of plane by regular octagons ($8$-gons) such that each lattice site has coordination number $3$. The hyperbolic lattice can be constructed via a recursive process (see Appendix~\ref{AppendixA})~\cite{PhysRevLett.125.053901}. In fact, our numerical calculations are on the finite hyperbolic $\{8,3\}$ lattice with open boundary condition. The finite hyperbolic $\{8,3\}$ lattice at the epoch $3$, represented by the Poincar\'{e} disk model, is shown in Fig.~\ref{fig1}, where the non-Euclidean metric is such that the distance between any two neighboring sites is equal.

Here, we will apply a 2D QAH model (Qi-Wu-Zhang model) proposed by Qi $et$ $al$~\cite{PhysRevB.74.085308} to the hyperbolic lattice. The Qi-Wu-Zhang model in the real space can be described by the following tight-binding model Hamiltonian:
\begin{align}
H_{0}=& (M-4t_{B})\sum_{j}c_{j}^{\dagger}\sigma_{z}c_{j}+t_{B}\sum_{\left< j,k \right>}c_{j}^{\dagger}\sigma_{z}c_{k}\nonumber \\
& +\frac{it_{A}}{2}\sum_{\left< j,k \right>}c_{j}^{\dagger}\left [ \cos(\theta_{jk})\sigma_{x}+\sin(\theta_{jk})\sigma_{y}\right ] c_{k},
\label{eq1}
\end{align}
where $c_{j}^{\dagger}$ and $c_{j}$ are the creation and annihilation operators of electrons on site $j$, $\theta_{jk}$ represents the polar angle of the vector from the site $k$ to the site $j$ in the Poincar\'{e} disk. $\sigma_{x,y,z}$ are the Pauli matrices representing orbital. $t_{A}$, $t_{B}$, and $M$ are material parameters. In the following calculations, we set $t_{A}=1$ and only consider coupling between the nearest neighbor sites in the Poincar\'{e} disk. Recently, several models have been shown to possess translation symmetry in hyperbolic lattices~\cite{boettcher2021crystallography, https://doi.org/10.48550/arxiv.2203.07292}, such as the hyperbolic Haldane model and hyperbolic Kane-Mele model. It should be noted that the Hamiltonian $H_0$ in Eq.~(\ref{eq1}) does not possess translation symmetry in the $\{8,3\}$ lattice in the presence of the polar angles $\theta_{jk}$, because $\theta_{jk}$ are defined in the Poincar\'{e} disk in the Euclidean plane, not projected from hyperbolic space.

\section{Numerical Results}
\label{Numerical Results}

\subsection{Clean limit}

\begin{figure}[t]
	\includegraphics[width=8.5cm]{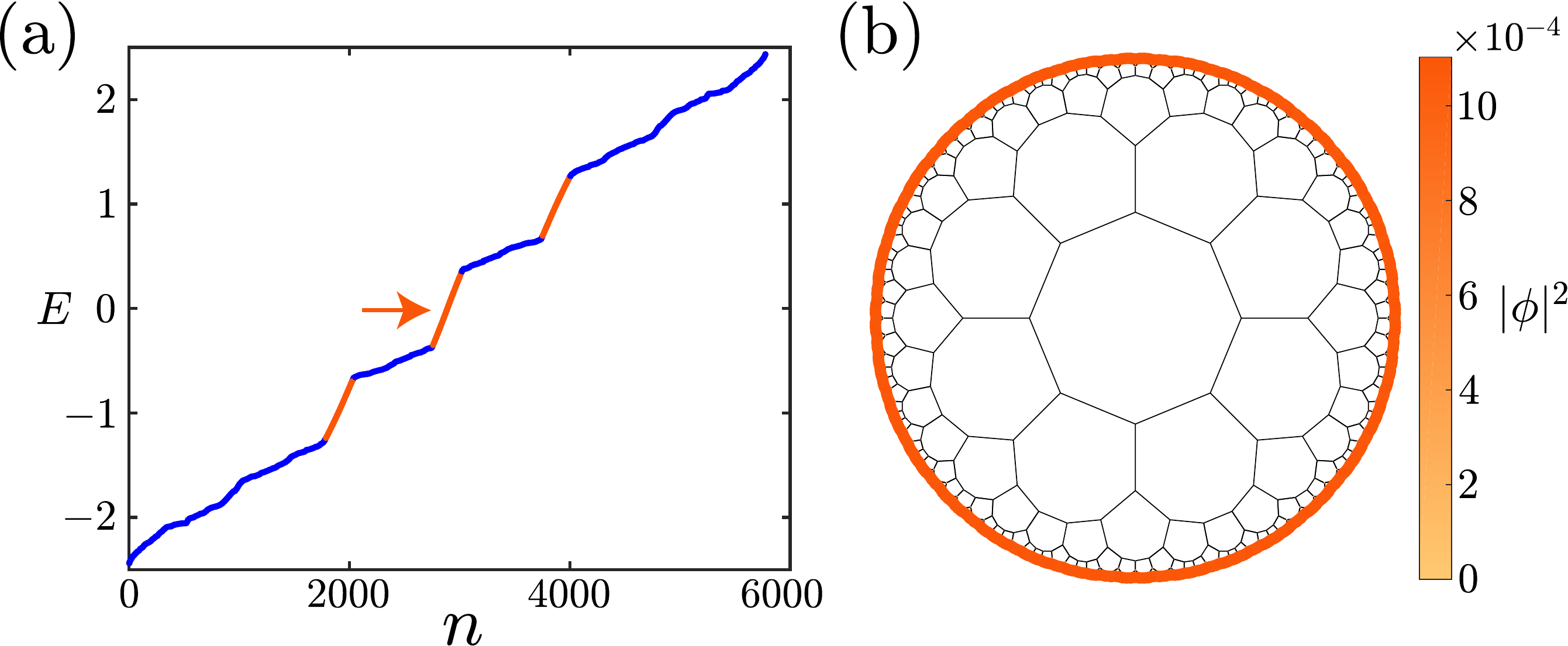} \caption{(a) Energy spectrum of the Hamiltonian $H_{0}$ on the hyperbolic $\{8,3\}$ lattice with open boundary condition. The three energy intervals marked by orange dots are $[-1.27, -0.67]$, $[-0.37, 0.37]$, and $[0.67, 1.27]$, respectively. The eigenstates of these energies are localized at the edge. In the main text, we mainly focus on the eigenstates near zero energy. (b) The probability distribution of the eigenstates near zero energy ($E_{n}=-0.0014$, $0.0014$) marked by the orange arrow in (a). We take the model parameters $t_{B}=0.5$, $M=1$ and the hyperbolic lattice site number $N=2888$.}%
	\label{fig2}
\end{figure}

First, in order to explore the electronic properties of the Qi-Wu-Zhang model in the hyperbolic lattice, we numerically diagonalize the Hamiltonian $H_{0}$ under open boundary condition. Figures~\ref{fig2}(a) and \ref{fig2}(b) show the energy spectrum of the system and the probability distribution of the eigenstate near zero energy, respectively. It can be found that the eigenstates near zero energy ($E_{n}=-0.0014$, $0.0014$) is localized at the edge of the hyperbolic lattice as shown in Fig.~\ref{fig2}(b). In fact, except that the eigenstates in the interval $E\in [-0.37, 0.37]$ near zero energy are localized at the edge, it is found that the eigenstates in the intervals $E\in [0.67, 1.27]$ and $E\in [-1.27, -0.67]$ are also localized at the edge. Here, we mainly focus on the eigenstates near zero energy, and the probability distribution of eigenstates in other energy intervals will be discussed in detail in Appendix \ref{AppendixC}. To analyze the topological property of the gapless edge states shown in Fig.~\ref{fig2}(b), we need to introduce a topological invariant to characterize the topological phase in the hyperbolic lattice described by the Hamiltonian $H_{0}$. It is well known that the Chern number is a useful topological invariant to characterize the CI topological phase. However, due to the absence of commutative translation symmetries and Bravais vectors, we cannot use the momentum-space Chern number to characterize the topological phase in hyperbolic lattices, and thus here we employ the real-space Bott index to characterize the topological phases of hyperbolic lattices.

The Bott index is equivalent to the Chern number~\cite{toniolo2017equivalence} and can be used to verify the topologically nontrivial phase of real-space systems with or without translation symmetries~\cite{Loring_2010, PhysRevLett.121.126401, PhysRevB.98.125130, PhysRevLett.116.257002, doi:10.1063/1.5083051, PhysRevB.100.085119, ghadimi2020topological, toniolo2017equivalence, PhysRevX.6.011016, PhysRevLett.118.236402, EXEL1991364, PhysRevB.104.155304}. To calculate the Bott index, firstly one constructs the projector operator of the occupied states as
\begin{align}
P=\sum_{n}^{N_{occ}}\left | \psi_{n} \right > \left < \psi_{n} \right |,
\end{align}
where $\left | \psi_{n} \right >$ is the $n$th eigenstate of the Hamiltonian with eigenvalue $E_{n}$ and $N_{occ}$ is the number of the occupied energy states. Here we set the eigenstate for the negative eigenvalue as the occupied energy state. Then, one defines the projected position operators as
\begin{align}
U=Pe^{i2\pi X}P,\\
V=Pe^{i2\pi Y}P,
\end{align}
where $X$ and $Y$ are diagonal matrices, and their diagonal elements $(x_{i},y_{i})$ are the rescaled coordinates of the $i$-th site in the Poincar\'{e} disk in the Euclidean plane, bounded in the interval $\left [0,1 \right )$. Finally, the Bott index is given by
\begin{align}
B=\frac{1}{2\pi} {\rm Im}\{{\rm tr}[{\rm log}(VUV^{\dagger}U^{\dagger})] \}.
\end{align}
A topologically nontrivial system is characterized by the Bott index $B=1$ (or $B=-1$), while the Bott index of a topological trivial system is equal to $0$. By calculating the Bott index, we confirm that the system with gapless edge states in Fig.~\ref{fig2}(b) corresponds to a topologically nontrivial phase (a CI phase with $B=1$).

\begin{figure}[t]
	\includegraphics[width=8.5cm]{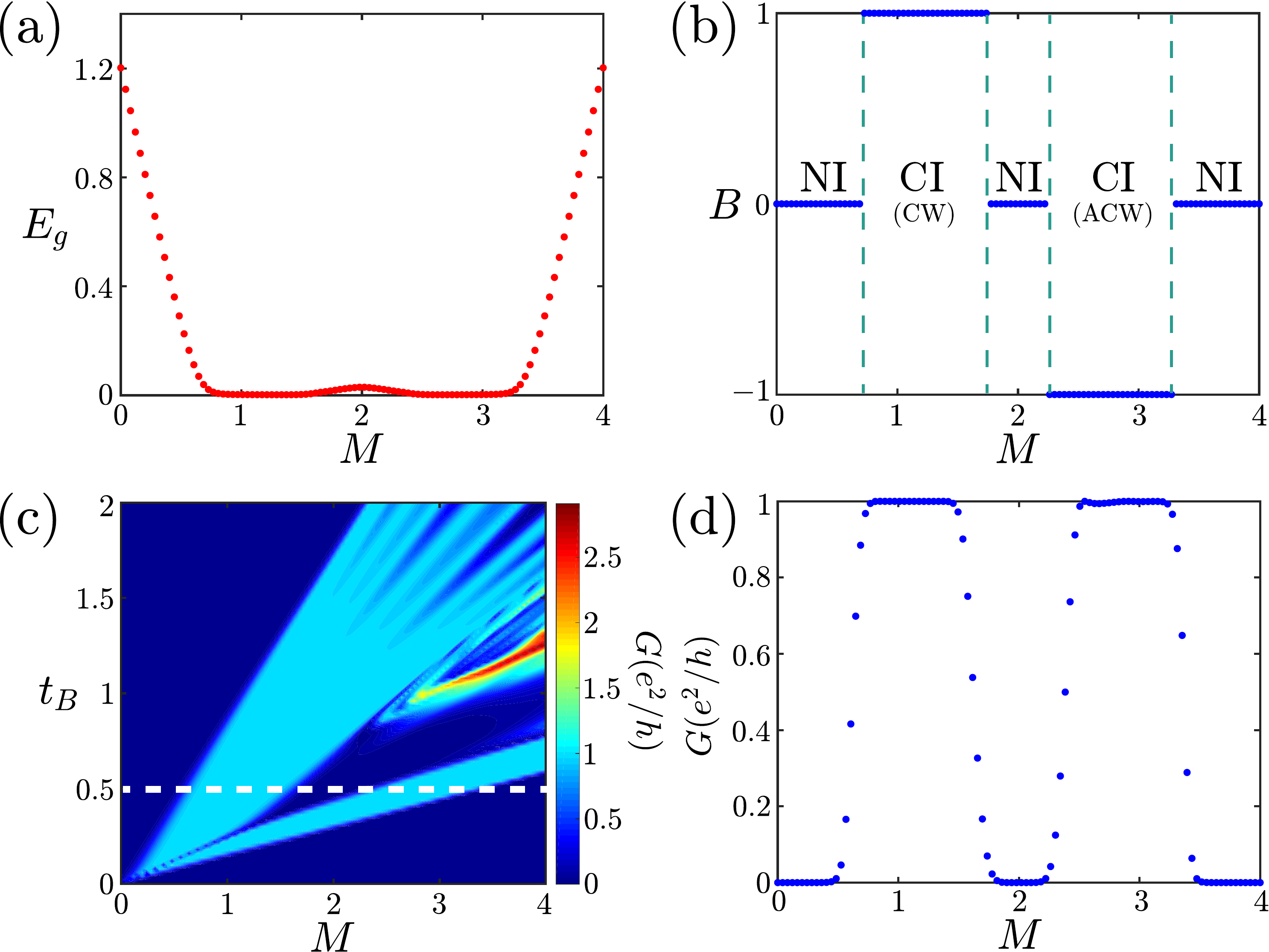} \caption{(a) Energy gap as a function of the parameter $M$. (b) Bott index as a function of $M$ under open boundary condition. (c) Phase diagram showing the conductance as a function of the model parameters $M$ and $t_{B}$. (d) Conductance as a function of $M$. We take the model parameters $t_{B}=0.5$ in (a), (b) and (d). Here, the hyperbolic lattice site number is $N=2888$.}%
	\label{fig3}
\end{figure}

In the Euclidean plane, it has been proved that the parameter $M$ can be used to adjust the topological phase of the system~\cite{PhysRevB.74.085308}. It is necessary to verify the adjustment effect of the parameter $M$ on the topological phase transition of hyperbolic lattices. Therefore, we calculate the energy gap of a finite hyperbolic $\{8,3\}$ lattice as a function of the parameter $M$ shown in Fig.~\ref{fig3}(a). It is found that a gapless phase can be obtained by adjusting the parameter $M$. To explore topological phase transition of the system when the parameter $M$ is adjusted, we calculate the Bott index of the hyperbolic lattice. In Fig.~\ref{fig3}(b), we present the Bott index with respect to $M$, and it is shown that the Bott index is $B=1$ ($B=-1$) in the parameter regimes where the gapless phases emerge. It means that the gapless edge state corresponds to a CI phase characterized by the nontrivial Bott index. Thus, we verify that the hyperbolic lattice described by the Hamiltonian $H_{0}$ can be transformed from a normal insulator (NI) to a CI, and the reverse is also true, by adjusting the parameter $M$. In fact, in addition to the Bott index, there are other topological invariants that can also characterize the topological properties of real-space systems, such as the real-space Chern number~\cite{KITAEV20062, PhysRevB.84.241106, https://doi.org/10.48550/arxiv.2203.07292, 10.1038/s41467-022-30631-x, Mitchell_2018}. In Appendix~\ref{AppendixB}, we further verify the above results by calculating the real-space Chern number.

In a 2D system, transport is a convenient way to experimentally verify the topologically nontrivial property of the system. To further confirm the topological states, we study the transport properties of the system. We design a two-terminal setup to calculate the conductance of the hyperbolic lattice. The setup consists of three parts, the central area is the hyperbolic lattice as the device, and the left and right ends of the device are connected with normal metal leads. Then we investigate the transport properties of the hyperbolic lattice based on the Landauer-B$\ddot{\text{u}}$ttiker-Fisher-Lee formula~\cite{doi:10.1080/14786437008238472, PhysRevB.38.9375, PhysRevB.23.6851} and the recursive Green's function method~\cite{MacKinnon_1985, PhysRevB.72.235304, PhysRevLett.127.066801, PhysRevResearch.3.033227, PhysRevB.100.205302}. The conductance can be obtained by
\begin{align}
G=\frac{e^{2}}{h}T,
\end{align}
where $T={\rm tr}[\Gamma_{L}G^{r}\Gamma_{R}G^{a}]$ is the transmission coefficient. The linewidth function $\Gamma_{l}=i[\Sigma_{l}^{r}-\Sigma_{l}^{a}]$ with $l=L,R$, and the retarded (advanced) Green's functions $G^{r(a)}$ of the device are calculated from $G^{r}=[G^{a}]^{\dagger}=\left [ \mu I-H_{D}-\Sigma_{L}^{r}-\Sigma_{R}^{r} \right ]^{-1}$, where $\mu$ is the chemical potential, $H_{D}$ is the device Hamiltonian, and $\Sigma_{L,R}^{r(a)}$ are the retarded (advanced) self-energies of the leads. We set the chemical potential of the device $\mu=0$ in the following calculations.

The conductance as a function of the parameter $M$ is plotted in Fig.~\ref{fig3}(d). Comparing Figs.~\ref{fig3}(b) and \ref{fig3}(d), it is found that the conductance of the region where the Bott index is equal to $1$ ($-1$) exhibits a quantized plateau with $G=e^{2}/h$. A slight difference of the phase boundaries between Figs.~\ref{fig3}(b) and \ref{fig3}(d) is owing to the finite size effect. Furthermore, the phase diagram of conductance as a function of $M$ and $t_{B}$ is presented in Fig.~\ref{fig3}(c), where the white horizontal dashed line is placed at $t_{B}=0.5$. The sky blue areas represent the topologically nontrivial phase with conductance $G=e^{2}/h$.

It is worth mentioning that two topologically nontrivial regions are found in Fig.~\ref{fig3}(b). The calculation result of the Bott index shows that these two regions represent two topological phases with opposite chirality. In order to further understand the transport properties of edge states in these two  topologically nontrivial regions, we calculate the nonequilibrium local current distribution of the hyperbolic lattice. A small external bias $V=V_{L}-V_{R}$ is applied between the two terminal, the local current from site \textbf{i} to site \textbf{j} can be calculated based on the formula~\cite{PhysRevB.76.153302, PhysRevB.78.155413, PhysRevB.80.165316, PhysRevB.100.115311, PhysRevB.103.085307, PhysRevB.103.L241409}:
\begin{align}
J_{\rm \textbf{i} \to \rm \textbf{j}}=\frac{2e^{2}}{h}{\rm Im}\left [\sum_{\alpha\beta}H_{\rm \textbf{i}\alpha,\rm \textbf{j}\beta}G^{n}_{\rm \textbf{j}\beta,\rm \textbf{i}\alpha}\right ](V_{L}-V_{R}),
\end{align}
where $H_{\rm \textbf{i}\alpha,\rm \textbf{j}\beta}$ is the matrix element of Hamiltonian $H_{D}$ of the device, and $\alpha$, $\beta$ represents orbital. $G^{n}=G^{r}\Gamma_{L}G^{a}$ is the electron correlation function.

\begin{figure}[t]
	\includegraphics[width=8cm]{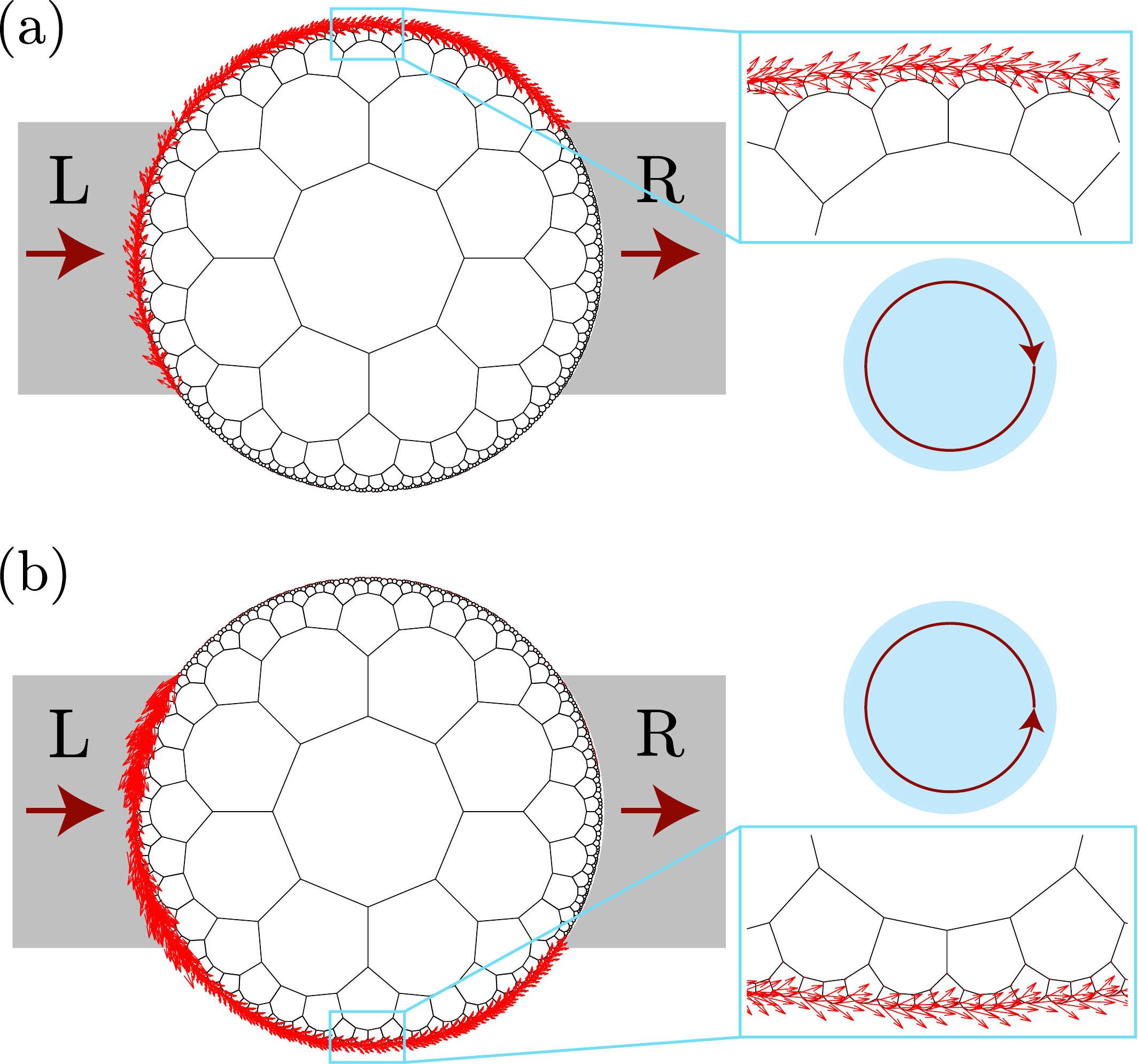} \caption{Nonequilibrium local current distribution of the CI phase (corresponding to $B=1$) and the CI phase (corresponding to $B=-1$) with $M=1$ for (a) and $M=3$ for (b). The gray shaded areas represent the leads, and the dark red arrow in the area indicates the direction of current injection and outflow. The middle disk is the hyperbolic lattice as the device, and the direction of the red arrow represents the direction of the local current, the arrow size means the local current strength. We take the model parameters $t_{B}=0.5$ and the hyperbolic lattice site number $N=2888$.}%
	\label{fig4}
\end{figure}

Figures~\ref{fig4}(a) and \ref{fig4}(b) show the local current distributions of the hyperbolic lattice with the parameters $M=1$ and $M=3$, respectively. In Fig.~\ref{fig4}, the middle disk is the hyperbolic lattice as the device, the left (L) and right (R) gray shaded parts denote the metal leads, and the dark red arrows in the shaded part indicate the direction of current input and output. The red arrow in the device represents the local current, and the size of the arrow means the strength of the local current. Obviously, the transport directions of electrons on the boundary of the two topologically nontrivial regions are exactly opposite, i.e., clockwise (CW) and anticlockwise (ACW). The directions of the arrows in the two light blue discs shown in Fig.~\ref{fig4} imply that the edge states in these two CI phases possess opposite chirality. Thus, the result of the local current distributions is consistent with that of the Bott index.

\subsection{Disorder Effect}

\begin{figure}[t]
	\includegraphics[width=8.5cm]{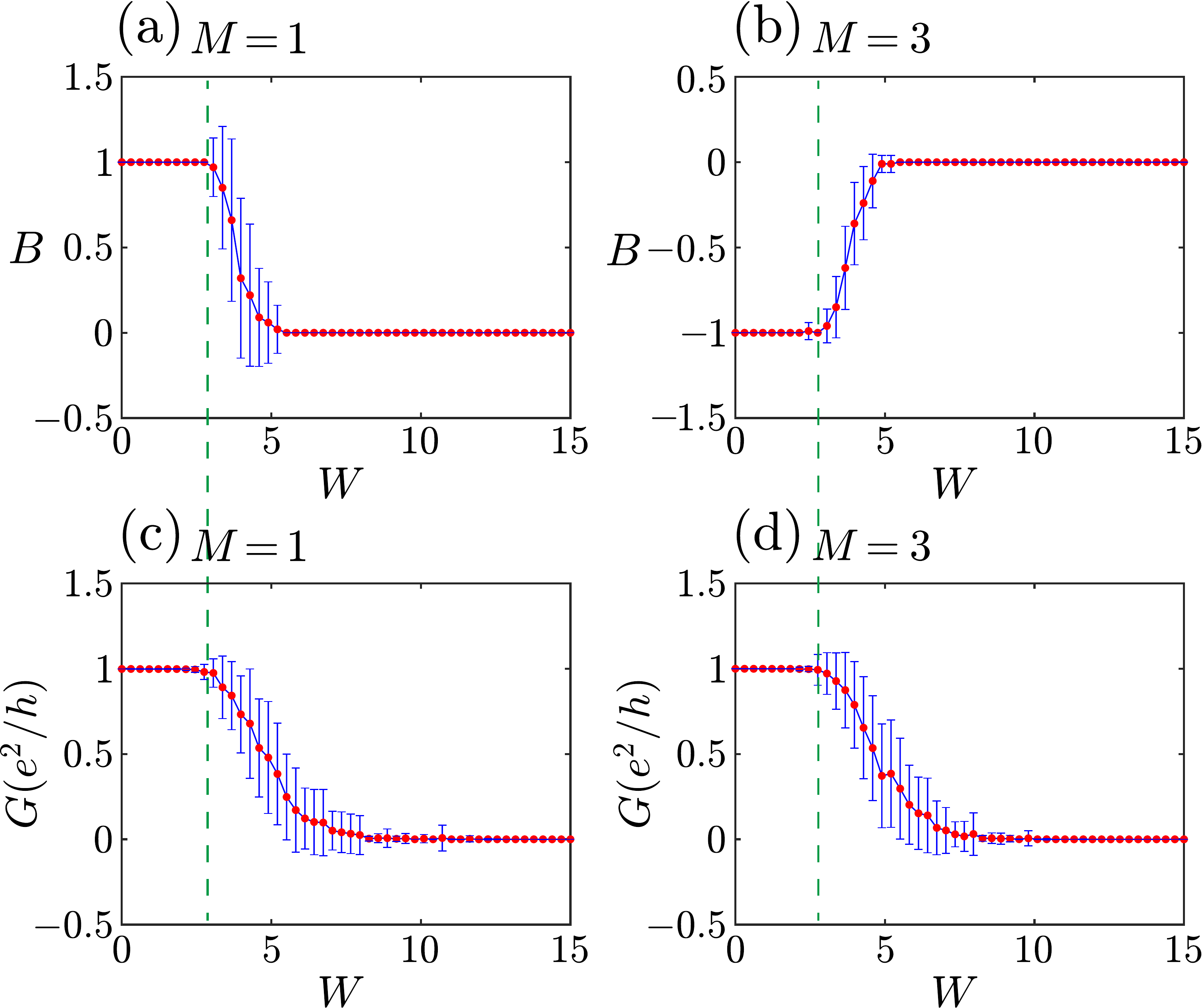} \caption{Bott index $B$ as a function of disorder strength $W$ with $M=1$ for (a) and $M=3$ for (b). (c) and (d) show the conductance $G$ as a function of disorder strength $W$ when $M=1$ and $M=3$, respectively. The error bar represents the standard deviation of 100 samples. We take the model parameters $t_{B}=0.5$ and the hyperbolic lattice site number $N=2888$.}%
	\label{fig5}
\end{figure}

Now, we focus on the effects of disorder on topological phases of the hyperbolic lattice system. We introduce on-site disorder to the system and corresponding Hamiltonian with the disorder is given by
\begin{align}
H_{\rm dis}=H_{0}+W\sum_{j}c_{j}^{\dagger}\omega_{j}\sigma_{0}c_{j},
\end{align}
where $W$ represents the disorder strength, and $\omega_{j}$ is a uniformly distributed variable within $[-0.5, 0.5]$. First of all, we identify the robustness of topologically nontrivial states against disorder. Therefore, we take two sets of parameters corresponding to the topologically nontrivial phases in the clean limit. For the case of $M=1$, in the clean limit, the system is a CI with the edge states transported in the CW direction. As shown in Fig.~\ref{fig5}(a), the topologically nontrivial phase with the Bott index $B=1$ remains stable within a certain range of disorder strength ($0\le W\le 2.8$). Beyond this range, the Bott index continues to decay as the disorder strength increases until the system transforms into a topologically trivial phase with $B=0$. Meanwhile, the conductance shown in Fig.~\ref{fig5}(c) remains a quantized conductance plateau under weak disorder. However, the strong disorder forces the quantized conductance to decrease until it converts to a localized state with conductance $G=0$. For the case of $M=3$, the chirality of the edge states of the system in the clean limit is opposite to that of $M=1$. Nevertheless, with increasing the disorder strength, the system with parameter $M=3$ shows a similar behavior, that is, both the Bott index and the conductance keep the quantized value until the disorder strength $W$ exceeds $2.8$ as shown in Figs.~\ref{fig5}(b) and \ref{fig5}(d).

\begin{figure}[t]
	\includegraphics[width=8.5cm]{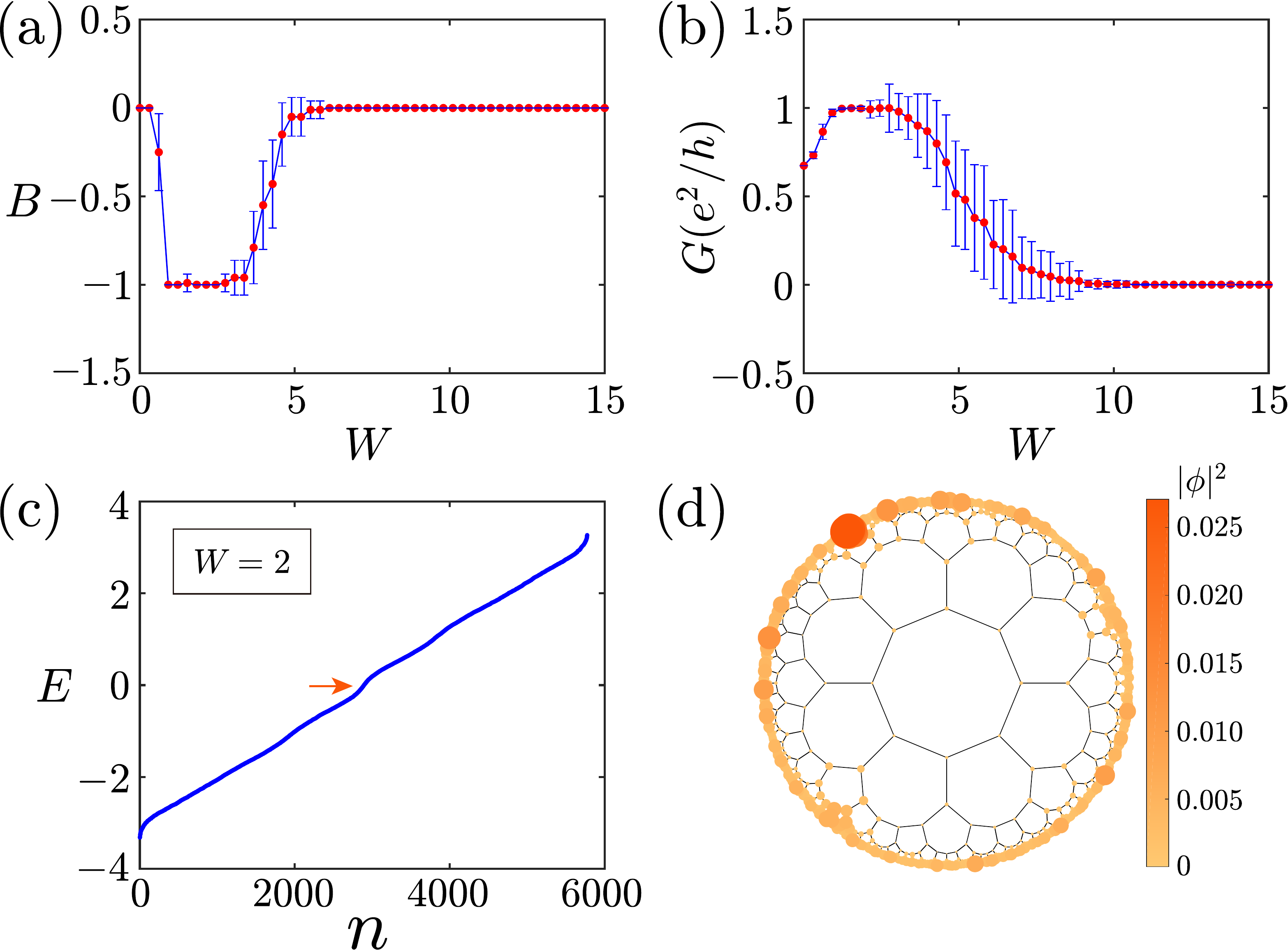} \caption{(a) Bott index $B$ and (b) the conductance $G$ as a function of disorder strength $W$ in the system with $M=3.3$. (c) Energy spectrum of a single sample with $M=3.3$ and the disorder strength $W=2$. (d) The probability distribution of the eigenstate near zero energy marked by the orange arrow in (c). The error bar represents the standard deviation of 100 samples. We take the model parameters $t_{B}=0.5$ and the hyperbolic lattice site number $N=2888$.}%
	\label{fig6}
\end{figure}

Next, the Bott index and the conductance with the model parameter $M=3.3$ as a function of disorder strength are plotted in Figs.~\ref{fig6}(a) and \ref{fig6}(b), respectively. In the clean limit [see Fig.~\ref{fig3}(b)], it is found that the system with $M=3.3$ is a NI phase characterized by the Bott index $B=0$. When the disorder is introduced, with the disorder strength increasing, it is observed from Fig.~\ref{fig6}(a) that the transition of the Bott index $B$ from $0$ to $-1$ occurs, and then the Bott index maintains at this value of $B=-1$ for a certain range before eventually goes back to the value of $B=0$. In Fig.~\ref{fig6}(b), it is found that with the disorder strength increasing, the conductance reaches a quantized value of $G=e^{2}/h$ and maintains this quantized plateau for a certain range of the disorder strength corresponding to the Bott index $B=-1$ shown in Fig.~\ref{fig6}(a). For the stronger disorder strength, the conductance gradually decreases to zero. The results shown in Figs.~\ref{fig6}(a) and \ref{fig6}(b) indicate that the disorder-induced topologically nontrivial phases can also occur in the hyperbolic lattice with disorder. Figure~\ref{fig6}(c) shows the energy spectrum of the system with $M=3.3$ and the disorder strength $W=2$. According to Fig.~\ref{fig6}(a), the eigenstate near zero energy marked by the orange arrow in Fig.~\ref{fig6}(c) corresponds to the disorder-induced CI phase with $B=-1$. The probability distribution of the eigenstate further confirms the disorder-induced CI phase supports localized edge states as shown in Fig.~\ref{fig6}(d).

\section{Conclusion and discussion}
\label{Conclusion}

In this work, we introduce the CI model into a hyperbolic $\{8,3\}$ lattice. Based on the numerical calculations of the Bott index and the two-terminal conductance, two CI phases characterize by $B=\pm1$ accompanied with quantized conductance plateaus are found in the hyperbolic lattice. We also plot the nonequilibrium local current distribution, which further confirms that the quantized conductance plateau originates from the chiral edge states and the two CI phases exhibit opposite chirality. Moreover, the effect of disorder on topological phases in the hyperbolic lattice is also analyzed. The robustness against the weak disorder of the chiral edge states of CIs in the hyperbolic lattice is confirmed. More fascinating, we observe a disorder-induced topological transition from a NI phase to a CI phase at a finite disorder strength. Our results present a non-Euclidean analog of TAI.

It is noted that the absence of commutative discrete translation groups limits most of theoretical research methods of hyperbolic lattices to real-space numerical diagonalization~\cite{Koll_r_2019, PhysRevLett.125.053901, bienias2021circuit, PhysRevA.102.032208}. In this work, all our numerical calculations are based on open boundary conditions. Recently, Maciejko and Rayan proposed a hyperbolic band theory based on the idea of algebraic geometry~\cite{doi:10.1126/sciadv.abe9170}. Subsequently, hyperbolic lattice band theory under a magnetic field and the crystallography of hyperbolic lattice have further been developed~\cite{Ikeda_2021, boettcher2021crystallography, doi:10.1073/pnas.2116869119}. Thus, research of topological states of matter in hyperbolic lattices based on periodic boundary conditions and hyperbolic topological band theory will be worth investigating in future works.

\section*{Acknowledgments}
B.Z. was supported by the NSFC (under Grant No. 12074107) and the program of outstanding young and middle-aged scientific and technological innovation team of colleges and universities in Hubei Province (under Grant No. T2020001).

\appendix

\section{Hyperbolic lattice construction}
\label{AppendixA}

\begin{figure}[h]
	\includegraphics[width=8cm]{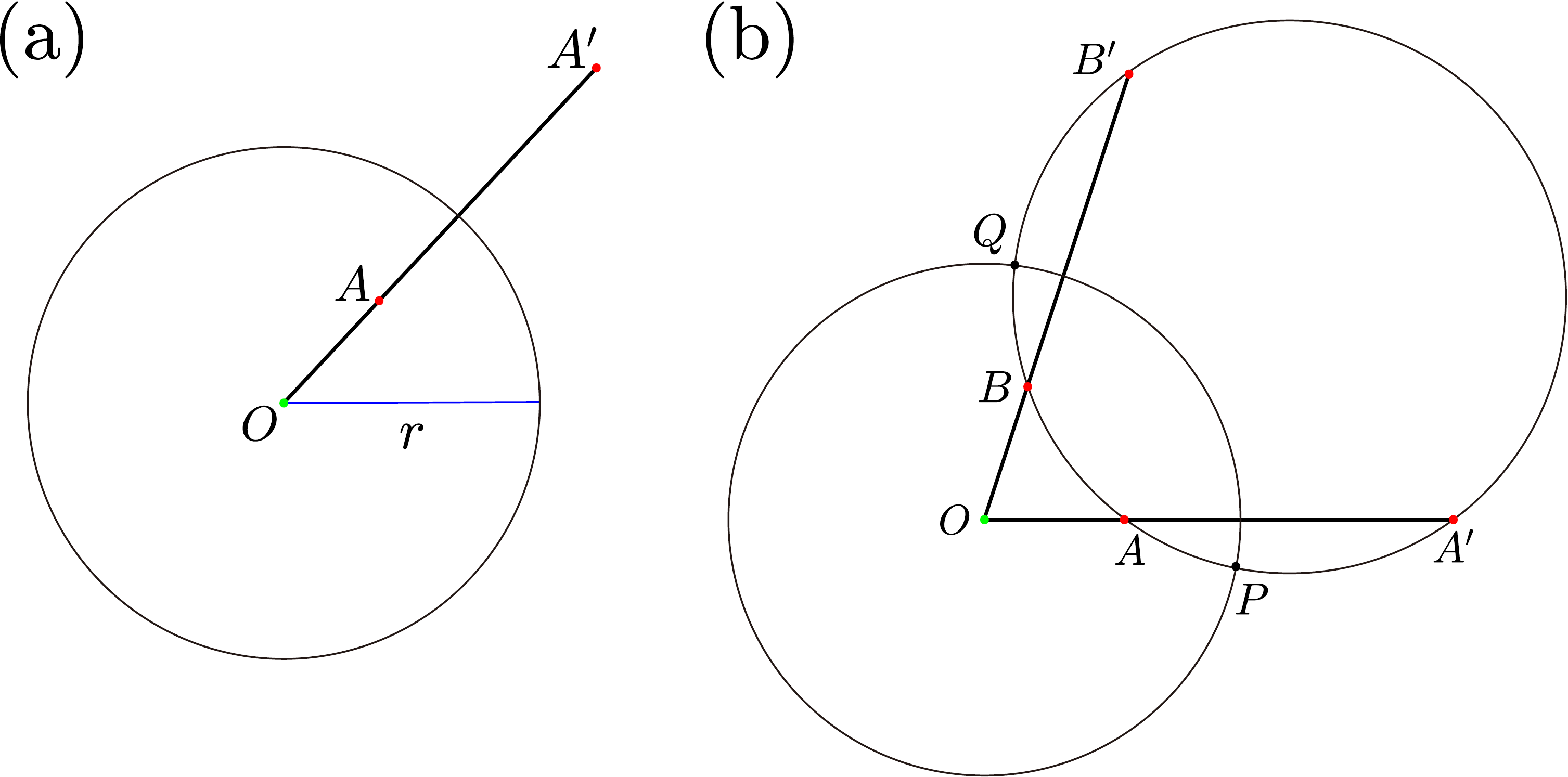} \caption{(a) The inversion circle $\textbf{c}$ has radius $r$ and center $O$, the initial point is $A$, and the inverted point is $A'$. (b) Starting points $A$ and $B$, inversion points $A'$ and $B'$ lie on a new circle, and the new circle intersects the inversion circle at points $P$ and $Q$.}%
	\label{fig7}
\end{figure}

In Appendix A, we present the details of hyperbolic lattice construction. Before we start constructing a hyperbolic lattice, let's understand the concept of circular inversion~\cite{Hartshorne_2000}. As shown in Fig.~\ref{fig7}(a), if the inversion circle $\textbf{c}$ has radius $r$ and center $O$, then the inverted point $A'$ of a point $A$ lies on a ray from $O$ to $A$, and
\begin{align}
OA \cdot OA' = r^{2}.
\end{align}

\begin{figure}[t]
	\includegraphics[width=8.5cm]{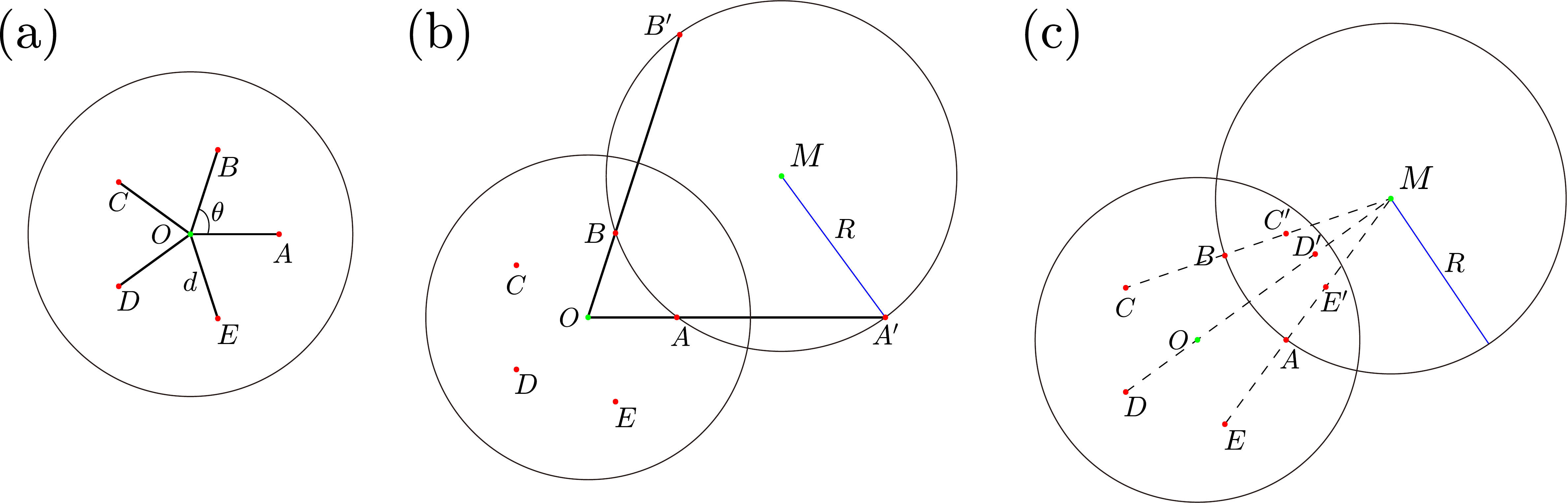} \caption{(a) Points $A$, $B$, $C$, $D$ and $E$ serve as the initial points of the hyperbolic lattice. (b) Points $A'$ and $B'$ are the inverted points of points $A$ and $B$ within circle $O$, which lie on the new circle $M$. The initial points $A$, $B$, $C$, $D$ and $E$ perform circle inversion for circle $M$, and the inversion points $A$, $B$, $C'$, $D'$ and $E'$ can be obtained in (c).}%
	\label{fig8}
\end{figure}

Unlike in the Euclidean plane, in the Poincar\'{e} disk the shortest distance between two points is not a straight line~\cite{weeks2001shape, greenberg1993euclidean, doi:10.1080/00029890.2001.11919719, petrunin2021euclidean}. Taking Fig.~\ref{fig7}(b) as an example, we use the following formula to calculate the distance between two points in the Poincar\'{e} disk~\cite{petrunin2021euclidean},
\begin{align}
d(A,B)=|\rm {ln}(A, B; P, Q)|=\left | \rm {ln}(\frac{AP\cdot BQ}{BP\cdot AQ}) \right |.
\end{align}

\begin{figure}[h]
	\includegraphics[width=8.5cm]{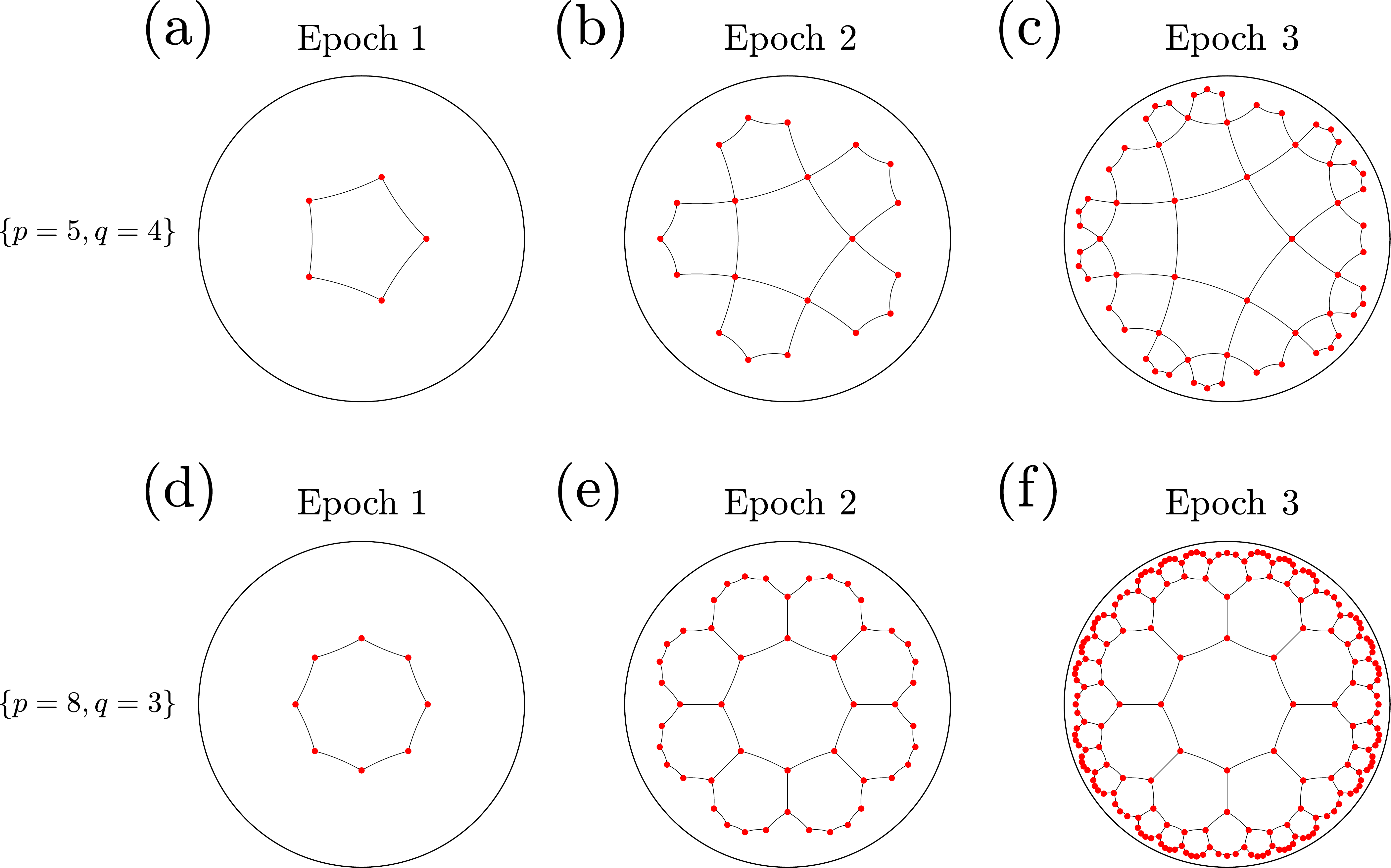} \caption{(a), (b) and (c) represent the site distribution of $\{p=5,~q=4\}$-type hyperbolic lattice at epochs 1, 2 and 3, respectively. (d), (e) and (f) represent the site distribution of $\{p=8,~q=3\}$-type hyperbolic lattice at epochs 1, 2 and 3, respectively.}%
	\label{fig9}
\end{figure}

Then, we describe the method of constructing a hyperbolic lattice by taking a hyperbolic lattice of $\{p=5,~q=4\}$-type as an example. As shown in Fig.~\ref{fig8}(a), if the center of the Poincar\'{e} disk is $O$, these $p$ points maintain the same distance $d$ from point $O$. This distance $d$ is
\begin{align}
d=\sqrt{\frac{\rm {tan}(\frac{\pi}{2}-\frac{\pi}{q})-\rm {tan}(\frac{\pi}{p})}{\rm {tan}(\frac{\pi}{2}-\frac{\pi}{q})+\rm {tan}(\frac{\pi}{p})}},
\end{align}
where $p$ and $q$ represent the number of vertices of the regular polygon and the number of polygons adjacent to each vertex, respectively. And we connect these $p$ points to the center of the circle $O$ with lines, and the angle between the lines corresponding to any two adjacent points is $\theta=2\pi/p$. Next, we perform circular inversion of circle $O$ for point $A$ and point $B$ respectively. We can confirm that the points $A'$ and $B'$ obtained by the circle inversion and the points $A$ and $B$ inside Circle $O$ form a new circle $M$, as shown in Fig.~\ref{fig8}(b). The radius of Circle $M$ is $R$. In Fig.~\ref{fig8}(c), we do a circular inversion of circle $M$ by Point $A$, $B$, $C$, $D$, and $E$. The new polygon consisting of points $A$, $B$, $C'$, $D'$, and $E'$ is exactly the same as the regular polygon consisting of points $A$, $B$, $C$, $D$, and $E$ in the Poincar\'{e} disk. When we keep repeating the steps of Figs.~\ref{fig8}(b) and \ref{fig8}(c), we can get hyperbolic lattices with different epochs. In Fig.~\ref{fig9}, we show the site distribution for different epochs of two types of hyperbolic lattices $\{p=5,~q=4\}$ and $\{p=8,~q=3\}$.

\section{Real-space Chern number}
\label{AppendixB}

\begin{figure}[h]
	\includegraphics[width=8.5cm]{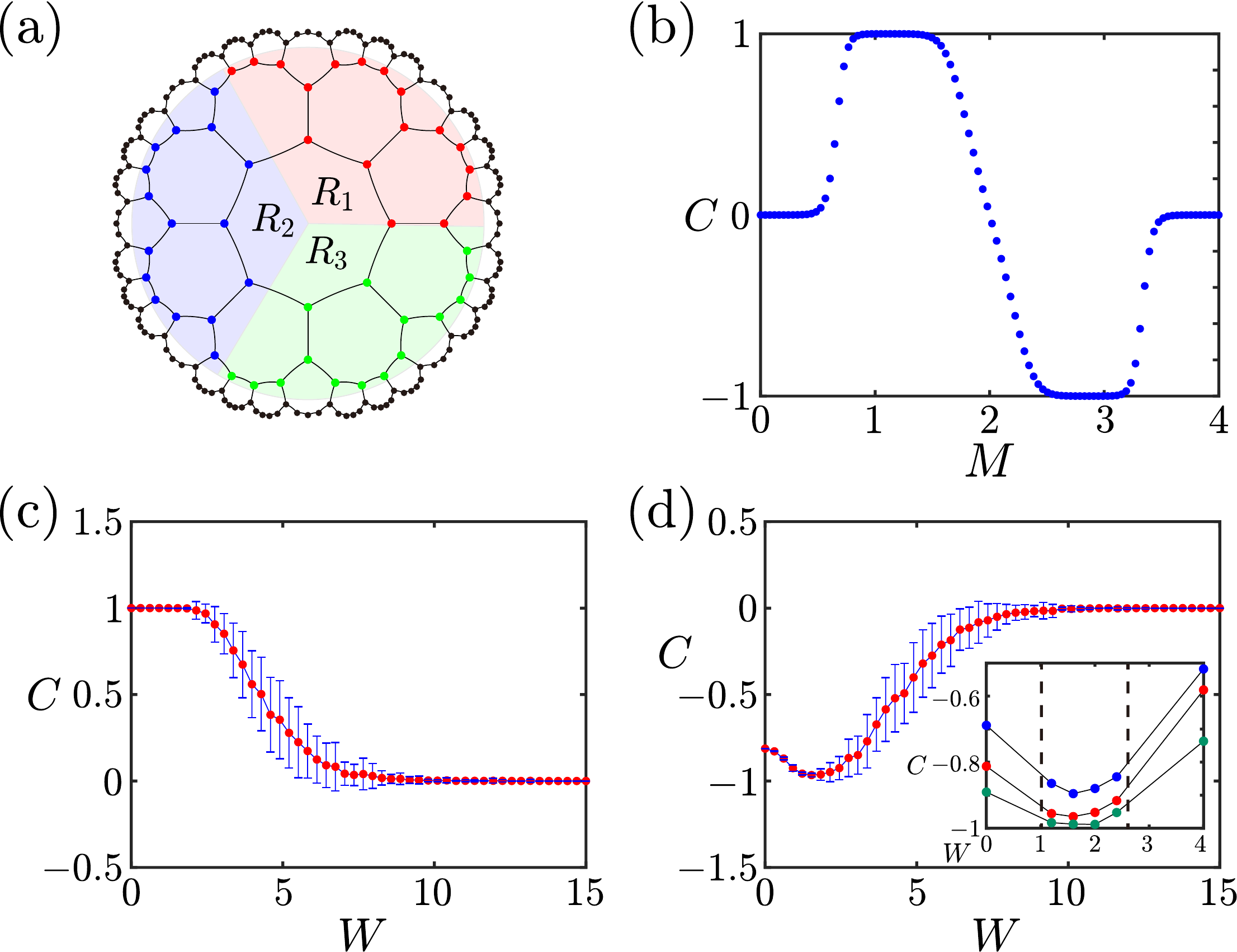} \caption{(a) The three regions $R_{1}$, $R_{2}$, and $R_{3}$ in the bulk of the system are arranged counterclockwise. (b) The real-space Chern number $C$ as a function of $M$. (c) The real-space Chern number $C$ as a function of disorder strength $W$ when $M=1$. The sample size used in (b) and (c) is $N=2888$. (d) The real-space Chern number $C$ as a function of disorder strength $W$ when $M=3.3$. The inset shows the real-space Chern number $C$ as a function of disorder strength $W$ for different sample sizes, with blue, red, and dark green points representing sample sizes $N=768$, $N=2888$, and $N=10800$, respectively. The error bar represents the standard deviation of 100 samples. We take the model parameter $t_{B}=0.5$ for (b), (c), and (d).}%
	\label{fig10}
\end{figure}

In Appendix B, we calculate another real-space topological invariant equivalent to the Bott index, the real-space Chern number. When the Fermi energy is $\epsilon$, the real-space Chern number is given by~\cite{KITAEV20062, PhysRevB.84.241106, https://doi.org/10.48550/arxiv.2203.07292, 10.1038/s41467-022-30631-x, Mitchell_2018},
\begin{align}
C(\epsilon)=12\pi i\sum_{j\in R_{1}}\sum_{k\in R_{2}}\sum_{l\in R_{3}}\left ( P_{jk}^{\epsilon}P_{kl}^{\epsilon}P_{lj}^{\epsilon}-P_{jl}^{\epsilon}P_{lk}^{\epsilon}P_{kj}^{\epsilon} \right ),
\end{align}
where $j$, $k$, and $l$ are site indices in three regions $R_{1}$, $R_{2}$, and $R_{3}$. These three regions only contain the bulk of the system and do not extend all the way to the boundary, and which are arranged counter-clockwise, as shown in Fig.~\ref{fig10}(a). $P^{\epsilon}$ is the projector operator of the occupied states taking the Fermi energy as $\epsilon$, $P^{\epsilon}=\sum_{E_{n} \leq \epsilon}\left | \psi_{n} \right > \left < \psi_{n} \right |$. $\psi_{n}$ is the eigenstate with the eigenvalue $E_{n}$. We set $\epsilon=0$ by default, unless the value of $\epsilon$ is specified. In comparison with Fig.~\ref{fig3}(b), we plot the real-space Chern number $C$ as a function of the parameter $M$ in Fig.~\ref{fig10}(b). It can be found that in the topologically nontrivial region of Bott index $B=1$ ($B=-1$), the real-space Chern number also exhibits a quantization platform of $C=1$ ($C=-1$). The differences near the phase boundaries are due to the finite size effect. In addition to clean systems, real-space Chern numbers can also characterize the topological properties of systems in disordered systems. Similar to Fig.~\ref{fig5}(a) and Fig.~\ref{fig6}(a), we plot the real-space Chern number $C$ as a function of disorder strength $W$ in Figs.~\ref{fig10}(c) and \ref{fig10}(d). It can be found that, like the Bott index, the real-space Chern number can also characterize the robustness of topologically nontrivial phase to weak disorder and the transition of system induced by disorder from a trivial phase to a nontrivial phase. Due to the finite size effect, the disorder-induced platform of $C=-1$ in Fig.~\ref{fig10}(d) is not quite perfect. To this end, we plot the variation of the platform (between the dashed black lines) at different sizes in the inset of Fig.~\ref{fig10}(d). It can be found that as the size increases, the platform gradually approaches $C=-1$ and becomes more flat.

\section{Energy spectrum and wave function}
\label{AppendixC}

\begin{figure}[h]
	\includegraphics[width=8.5cm]{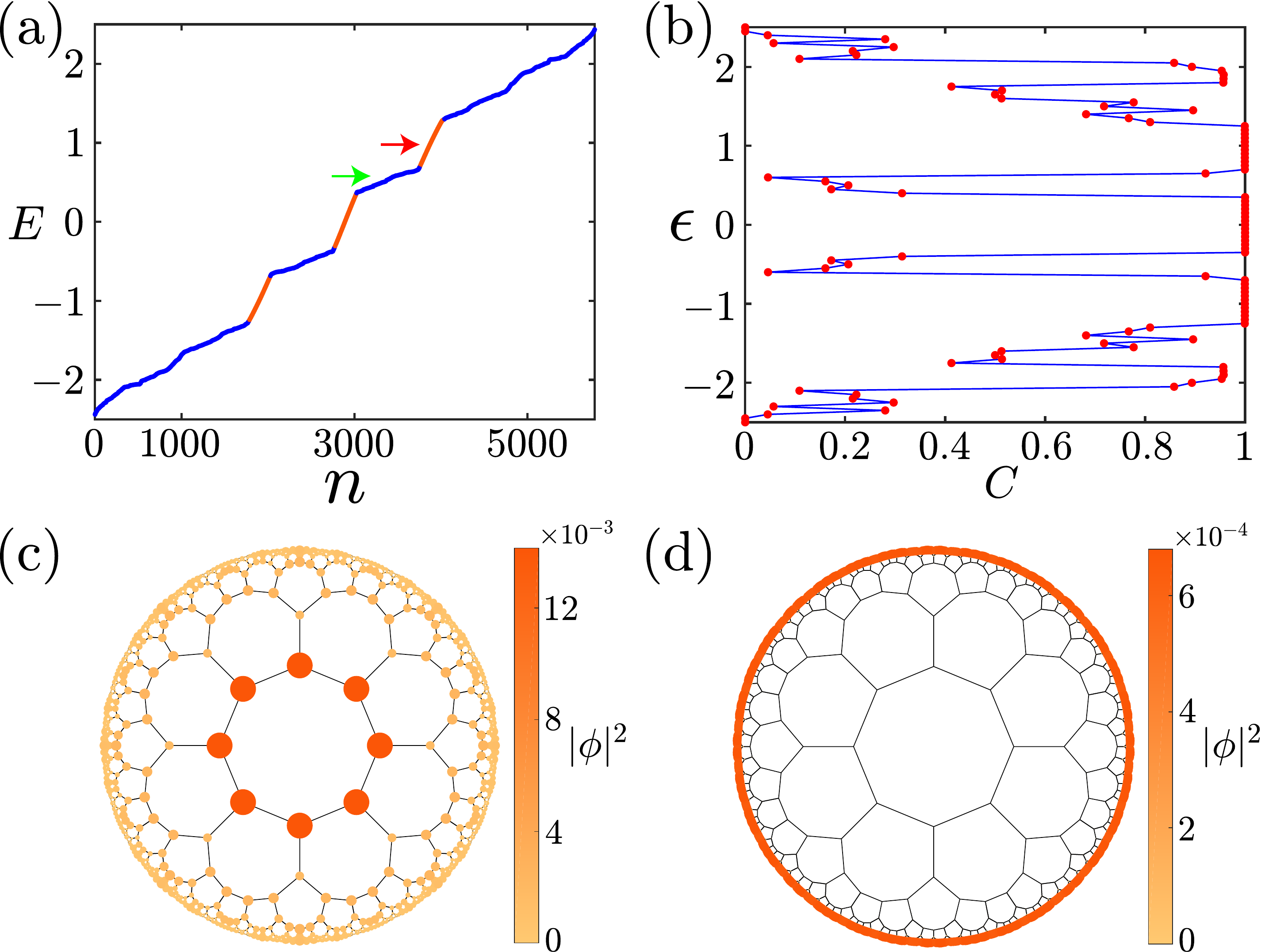} \caption{(a) Energy spectrum of the Hamiltonian $H_{0}$ on the hyperbolic $\{8,3\}$ lattice with open boundary condition. The three energy intervals marked by orange dots are $[-1.27, -0.67]$, $[-0.37, 0.37]$, and $[0.67, 1.27]$, respectively. The eigenstates of these energies are localized at the edge. (b) The real-space Chern number $C$ as a function of Fermi energy $\epsilon$. (c) The probability distribution of the eigenstate with the energy $E=0.5227$ marked by the green arrow in (a). (d) The probability distribution of the eigenstate with the energy $E=1.001$ marked by the red arrow in (a). Here, we take the model parameters $t_{B}=0.5$, $M=1$, and $N=2888$.}%
	\label{fig11}
\end{figure}

In Appendix C, we show the energy spectrum for different parameters, as well as the wave function distribution with different energies. In the clean limit, we show the energy spectrum for the three sets of parameters and the wave function distribution in different energy intervals. In the main text, we mentioned that when the parameters are taken as $t_{B}=0.5$ and $M=1$, in addition to the eigenstates near zero energy being localized at the edge, there are also some eigenstates with higher energies that are also localized at the edge. In Fig.~\ref{fig11}(a), we plot the energy spectrum when the parameters are taken as $t_{B}=0.5$ and $M=1$. We mark the energies of the eigenstates localized at the edge with orange dots, including three intervals, $[-1.27, -0.67]$, $[-0.37, 0.37]$, and $[0.67, 1.27]$. Furthermore, we plot the real-space Chern number $C$ as a function of the Fermi energy $\epsilon$ in Fig.~\ref{fig11}(b). It is found that in addition to the interval $[-0.37, 0.37]$, when the Fermi energy is placed in the interval $[-1.27, -0.67]$ and the interval $[0.67, 1.27]$, the real space Chern number is also equal to $1$, as shown in Fig.~\ref{fig11}(b). In the main text, we plot the probability distribution of near zero energy edge states in Fig.~\ref{fig2}(b). Here, in Fig.~\ref{fig11}(c) we plot the probability distribution of the eigenstate of the energy $E=0.5227$ marked with the green arrow in Fig.~\ref{fig11}(a), the real-space Chern number is not an integer when the Fermi energy is placed at this energy. Moreover, in Fig.~\ref{fig11}(d) we plot the probability distribution of the eigenstate of the energy $E=1.001$ marked with the red arrow in Fig.~\ref{fig11}(a), the real-space Chern number $C=1$ when the Fermi energy is placed at this energy.

\begin{figure}[h]
	\includegraphics[width=8.5cm]{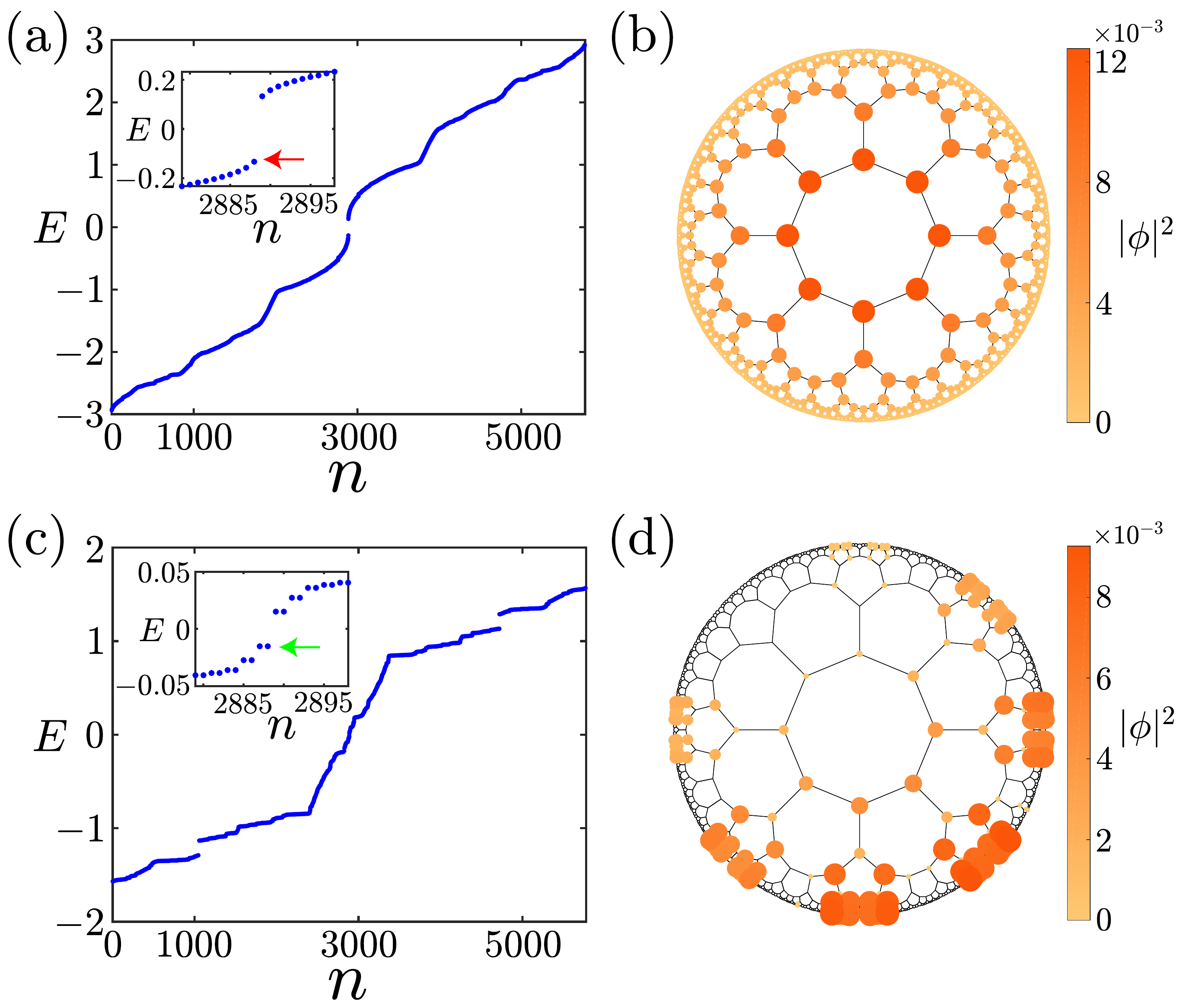} \caption{(a) The energy spectrum when we take parameter $M=0.5$. (b) The probability distribution of the eigenstate with the energy $E=-0.1332$ that is marked by the red arrow in (a). (c) The energy spectrum when we take parameter $M=2$. (d) The probability distribution of the eigenstate with the energy $E=-0.0151$ that is marked by the green arrow in (c). Here, we take the model parameters $t_{B}=0.5$ and $N=2888$.}%
	\label{fig12}
\end{figure}

In addition to the energy spectrum in the topologically nontrivial phase, we also analyze the energy spectrum in the NI phase and the wave function distribution of the partial energy. In Fig.~\ref{fig12}(a) we plot the energy spectrum of the NI phase with a large energy gap, and we plot the probability distribution of the eigenstate with the energy $E=-0.1332$ in Fig.~\ref{fig12}(b). Moreover, we also plot the energy spectrum of the NI phase between two Chern insulator phases with different chirality in Fig.~\ref{fig12}(c), and we plot the probability distribution of the eigenstate with the energy $E=-0.0151$ in Fig.~\ref{fig12}(d). As shown in Figs.~\ref{fig12}(a) and \ref{fig12}(c), there is an energy gap in the energy spectrum of both NI phases, while the energy gap of NI between the two CI phases is smaller, which is consistent with Fig.~\ref{fig3}(a) in the main text. In Figs.~\ref{fig12}(b) and \ref{fig12}(d), it can be found that the eigenstates are non-locally distributed in the bulk of the system.

\begin{figure}[t]
	\includegraphics[width=8.5cm]{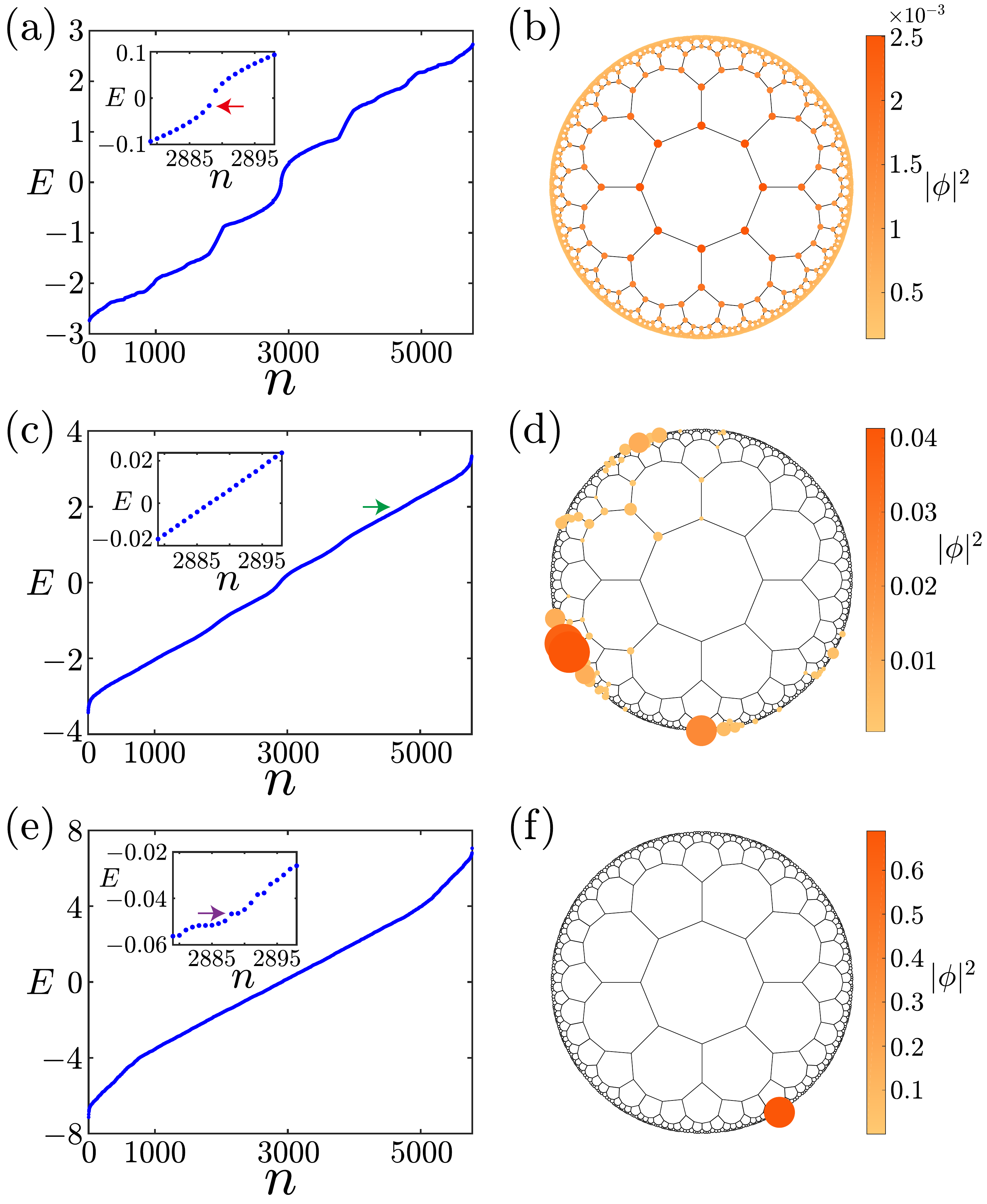} \caption{(a) The energy spectrum when we take parameter $M=3.3$. (b) The probability distribution of the eigenstate with the energy $E=-0.0163$ that is marked by the red arrow in (a). (c) The energy spectrum when we take parameters $M=3.3$ and $W=2$. (d) The probability distribution of the eigenstate with the energy $E=1.9912$ that is marked by the dark green arrow in (c). (e) The energy spectrum when we take parameters $M=3.3$ and $W=10$. (f) The probability distribution of the eigenstate with the energy $E=-0.0467$ that is marked by the purple arrow in (e). Here, we take the model parameters $t_{B}=0.5$ and $N=2888$.}%
	\label{fig13}
\end{figure}

Next, we show the energy spectrum and wave function distribution in the topologically trivial phase at the clean limit, however when we add disorder of appropriate strength, the system exhibits nontrivial properties. In clean limit, when we take the parameters $t_{B}=0.5$ and $M=3.3$, the system is in the trivial phase. As shown in Fig.~\ref{fig13}(a), there is a small energy gap in the energy spectrum. In Fig.~\ref{fig13}(b), we select an energy that is closest to zero, and plot the probability distribution of the eigenstate of this energy. It is found that it is delocalized and distributed in the bulk of the system. We already know from the main text that disorder with suitable strength can induce a topologically nontrivial phase in the system. In Fig.~\ref{fig6}(d), we plot the probability distribution of disorder-induced gapless edge states. Here, we plot the energy spectrum for parameters $t_{B}=0.5$, $M=3.3$, $W=2$ and the probability distribution of eigenstate with higher energy ($E=1.9912$) in Figs.~\ref{fig13}(c) and \ref{fig13}(d), respectively. It can be found that the eigenstate of this energy is localized, but unlike the nontrivial edge states, it is only localized in part of the system, which is the Anderson localization~\cite{PhysRev.109.1492, PhysRevLett.47.882, PhysRevLett.102.136806, RevModPhys.80.1355, PhysRevLett.42.673}.
When the disorder strength increases to $W=10$, the system is in the trivial phase. We plot the energy spectrum with parameters $t_{B}=0.5$, $M=3.3$, $W=10$ in Fig.~\ref{fig13}(e). Although the energy spectrum is also gapless at this time, it is found that the energy state exhibits Anderson localization as shown in Fig.~\ref{fig13}(f). The system does not possess quantized conductance at this time, and such a localized state is not a topologically nontrivial edge state.

\end{document}